\documentstyle[12pt]{article}

\begin{document}

\newcommand{\beq}{\begin{equation}}
\newcommand{\eeq}{\end{equation}}
\newcommand{\beqa}{\begin{eqnarray}}
\newcommand{\eeqa}{\end{eqnarray}}

\def\ov{\overline}
\def\onlyif{\rightarrow}

\def\openone{\leavevmode\hbox{\small1\kern-3.8pt\normalsize1}}

\def\a{\alpha}
\def\b{\beta}
\def\g{\gamma}
\def\r{\rho}
\def\minus{\,-\,}
\def\eks{\bf x}
\def\kay{\bf k}

\def\ket#1{|\,#1\,\rangle}
\def\bra#1{\langle\, #1\,|}
\def\braket#1#2{\langle\, #1\,|\,#2\,\rangle}
\def\proj#1#2{\ket{#1}\bra{#2}}
\def\expect#1{\langle\, #1\, \rangle}
\def\trialexpect#1{\expect#1_{\rm trial}}
\def\ensemblexpect#1{\expect#1_{\rm ensemble}}
\def\kpsi{\ket{\psi}}
\def\kphi{\ket{\phi}}
\def\bpsi{\bra{\psi}}
\def\bphi{\bra{\phi}}

\def\ditto{\rule[0.5ex]{2cm}{.4pt}\enspace}
\def\th{\thinspace}
\def\ni{\noindent}
\def\thirty{\hbox to \hsize{\hfill\rule[5pt]{2.5cm}{0.5pt}\hfill}}

\def\set#1{\{ #1\}}
\def\setbuilder#1#2{\{ #1:\; #2\}}
\def\Prob#1{{\rm Prob}(#1)}
\def\pair#1#2{\langle #1,#2\rangle}
\def\Id{\bf 1}

\def\dee#1#2{\frac{\partial #1}{\partial #2}}
\def\deetwo#1#2{\frac{\partial\,^2 #1}{\partial #2^2}}
\def\deethree#1#2{\frac{\partial\,^3 #1}{\partial #2^3}}

\newcommand{\xx}{{\scriptstyle -}\hspace{-.5pt}x}
\newcommand{\yy}{{\scriptstyle -}\hspace{-.5pt}y}
\newcommand{\zz}{{\scriptstyle -}\hspace{-.5pt}z}
\newcommand{\kk}{{\scriptstyle -}\hspace{-.5pt}k}
\newcommand{\sx}{{\scriptscriptstyle -}\hspace{-.5pt}x}
\newcommand{\sy}{{\scriptscriptstyle -}\hspace{-.5pt}y}
\newcommand{\sz}{{\scriptscriptstyle -}\hspace{-.5pt}z}
\newcommand{\sk}{{\scriptscriptstyle -}\hspace{-.5pt}k}

\def\openone{\leavevmode\hbox{\small1\kern-3.8pt\normalsize1}}

\title{Bell inequality for quNits with binary measurements}
\author{
H. Bechmann-Pasquinucci and N. Gisin
\\
\small
{\it Group of Applied Physics, University of Geneva, CH-1211, Geneva 4,
Switzerland}}
\date{April 21, 2002}
\maketitle

\abstract{We present a generalized Bell inequality for two entangled 
quNits. On one quNit the choice is between two standard von Neumann 
measurements, whereas for the other quNit there are $N^2$ different binary 
measurements. These binary measurements are related to the 
intermediate states known from eavesdropping in quantum 
cryptography. The maximum violation by $\sqrt{N}$  
is reached for the maximally entangled state. Moreover, for $N=2$ it 
coincides with the 
familiar CHSH-inequality.} 
\vspace{1 cm} \normalsize
\section{Introduction}
Recently there has been an increasing interest in generalizing results 
known for qubits to arbitrary dimensions. In this respect both quantum 
cryptography \cite{BT,Sw} and  various types of Bell 
inequalities \cite{bell}, \cite{bb}-\cite{bf} have been 
generalized. Here we combine the two, in the sense 
that we present 
a Bell inequality for two quNits ($N$ dimensional quantum systems), where 
the basic idea of the 
construction 
is inspired by quantum 
cryptography.

In the quantum cryptographic protocol, known as BB84 \cite{BB84}, the 
legitimate users Alice and Bob, both  chose between measuring in one of 
the two mutually unbiased bases $A$ and $A'$. However, an 
eavesdropper performing intercept/resend eavesdropping may chose to 
perform her measurements in what is known as the intermediate 
basis or the Breidbart basis \cite{exp}. In two dimensions it is possible 
to form two 
intermediate bases, but the eavesdropper needs only to make use of one of 
them. Turning to the Clauser-Horne-Shimony-Holt-inequality 
(CHSH) \cite{CHSH} for two entangled qubits. 
The 
maximal
violation is obtained when on the first qubit the measurement settings are
$A$ and $A'$, and on the second qubit the two intermediate bases.

It is this observation which lies at the heart of the construction of the 
inequality we present here. The 
intermediate states may be generalized to any dimension. However, 
in higher dimension the intermediate states do in general not form bases. 
But the projectors
corresponding to these intermediate states can be used as binary
measurements. 

This idea leads to an inequality for two entangled quNits, where on 
the first quNit the choice of measurement is between two mutually 
unbiased bases $A$ and $A'$, but on the second quNit the choice is 
between $N^2$ mutually incompatible binary measurements. These 
measurements 
correspond to the projectors of all  the possible intermediate states.
We find that the limit from a local variable point of view is $2$, whereas 
the quantum mechanical limit is $2\sqrt{N}$.

It should be emphasized that, the inequality we present here differs in 
various aspects from the ones 
which have recently been presented in the literature. First of all the 
choice of measurements: Usually it is assumed that Alice and Bob each have 
two measurement settings. Here Alice has again two, but Bob has $N^2$. 
Second, due to the special choice of measurements the construction of the 
inequality is easily generalized to any dimension. And finally, but very 
important, this inequality has contrary to other inequalities in higher 
dimensions, maximal violation for the maximally entangled state. This 
means that this inequality may be used as a measure of entanglement.

We show the full construction of the 
inequality for two qutrits and shortly discuss the generalization to 
arbitrary dimension. In section 2, we define the intermediate states for 
qutrits and in section 3 we obtain the corresponding inequality. In 
section 4 we show how to extend this result to any dimension. And then  
since the 
strength of a Bell inequality is often measured in terms of its reststance 
to noise, in section 5 we shortly discuss this issue for the inequality 
we have obtained. In section 6 we have the conclusions.

\section{The intermediate states}
In this and the next section we consider qutrits, i.e. $N=3$.

The Bell inequality we are about to present is derived from the 
measurement 
settings,  therefore we first define all the measurements involved.  
The measurement setting on one side, let's say the side of Alice, 
correspond to two mutually unbiased basis $A$ and $A'$ and on the side of 
Bob, the measurement settings correspond to all the intermediate states 
which may be constructed from these two bases. 
Here the  $A$-basis is chosen as the computational basis,
\beq \ket{a_0},\ket{a_1},\ket{a_2} \eeq
where the states satisfy $\braket{a_k}{a_l}=\delta_{kl}$.
The $A'$-basis is chosen as the Fourier transform of the computational
basis, i.e.
\beq \ket{{a}_{k}'}={\frac{1}{\sqrt{3}}}\sum_{n=0}^{2} \exp{\left(
\frac{2\pi i~
kn}{3}\right)}\ket{a_n}, \eeq
again these states satisfy
\beq
\braket{{a}_{k}'}{{a}_{l}'}=\delta_{kl},~~ {\rm and ~moreover}~~
\braket{a_k}{{a}_{l}'}=\frac{\exp(i\phi_{kl})}{\sqrt(3)},
\eeq
which means that the two bases are mutually unbiased, and that the
distance between any states from the two different bases is
$\cos(\theta)=1/\sqrt(3)$.

The intermediate states are obtained by forming all possible pairs 
of states from the two bases.  They are  shown in the table below
\begin{center}
\begin{tabular}{|l||l|l|l|} \hline
    & ${a}_{0}'$& ${a}_{1}'$& ${a}_{2}'$ \\ \hline \hline
$a_0$ & $m_{00}$ & $m_{01}$ &$m_{02}$ \\ \hline
$a_1$ & $m_{10}$ & $m_{11}$ &$m_{12}$ \\ \hline
$a_2$ & $m_{20}$ & $m_{21}$ &$m_{22}$ \\ \hline
\end{tabular}
\end{center}
where  $m_{ij}$ is understood as the intermediate state between the
states
$\ket{{a_i}}$, $\ket{{a}_{j}'}$, the first index always refers to 
the state from $A$  and the second to the state from $A'$. 

In quantum cryptography an  eavesdropper, performing the simple 
intercept/resend eavesdropping strategy, may use the intermediate 
states to make a guess of the identity of each state send by Alice. Since 
the eavesdropper learns the basis in which the particle was 
originally prepared, she uses the intermediate states in the 
following way: suppose 
the eavesdropper in a measurement finds the state $\ket{{m}_{ij}}$,  
if she subsequently learns that the basis was $A$, she concludes 
that most 
probably the 
original state was $\ket{{a_i}}$, whereas if she learns that the basis was 
$A'$, she will guess that most probably the state was 
$\ket{{a}_{j}'}$. This means that she wants to optimize the conditional 
probability 
\begin{equation}
p(m_{ij}|a_i)=  p(m_{ij}|{a}_{j}') = {\rm ~max~value}.
\end{equation}  
In other words she wants to optimize her probability for guessing the 
state correctly --- independently of the basis. But at the same time she 
also 
wants the errors to be 
evenly distributed between the 
wrong states, i.e.
\begin{equation}
p(m_{ij}|a_k)=  p(m_{ij}|{a}_{l}') ~~~k\neq i ~{\rm
and}~l\neq j.
\end{equation}
The intermediate state, $\ket{m_{ij}}$, fulfill these requirements 
\cite{BPG}. In terms of the two basis
states $\ket{a_{i}}$ and $\ket{{a}_{j}'}$, it can be written as  
\beq
\ket{m_{ij}}={\frac{1}{\sqrt{C}}}\left(
\exp{( i\phi_{ij})}\ket{a_i} +\ket{{a}_{j}'}\right)
\eeq
where $C=2(1+1/\sqrt{3})$ is the normalization constant, and the phase
comes from the overlap between $\ket{{a_i}}$ and $\ket{{a}_{j}'}$. This 
leads to the conditional probability 
\beq
p({m_{ij}}|a_i)=p({m_{ij}}|{a}_{j}')= \frac{1+\frac{1}{\sqrt{3}}}{2}=
\frac{1}{2}+\frac{1}{2\sqrt{3}}
\eeq
This can also be recognized as the cosine squared of half the angle, i.e.
${\cos}^{2}(\theta/2)=\frac{1+\cos(\theta)}{2}$. Which indeed shows that 
the 
intermediate state $\ket{{m_{ij}}}$, is as the name indicates, lying 
exactly between  the states
$\ket{{a_j}}$ and  
$\ket{{a}_{j}'}$. The probability for obtaining a wrong state 
is
\beq p({m_{ij}}|{a_k})=p({m_{ij}}|{a}_{l}')=\frac{1}{2}
\left(\frac{1}{2}-\frac{1}{2\sqrt{3}}\right)
\eeq
In this way the total probability for making an error is
$\frac{1}{2}-\frac{1}{2\sqrt{3}}$.

Notice that there has been made no requirement for orthogonality and 
indeed it may be checked that none of the nine states $\ket{m_{ij}}$ are
orthogonal\footnote{However, it turns out that the nine states constitute 
a generalized measurement namely a so called POVM. We have,
$
\sum_{k,l=0}^{2}\frac{1}{3}\ket{m_{ij}}\bra{m_{ij}}=\openone.
$}.
However, each of the nine 
states
can be associated with a projector 
$\ket{m_{ij}}\bra{m_{ij}}$, which may be identified as a binary 
measurements. These nine mutually incompatible binary measurements are the 
measurement settings 
on Bob side.

\section{The Bell inequality}
Assume that the two observers Alice and Bob share many maximally entangled 
state of two 
qutrits. In the two bases $A$ and $A'$ this state may be written as
\begin{eqnarray}
\ket{\psi}&=&\frac{1}{\sqrt{3}}\left(\ket{a_0,a_0}+
               \ket{a_1,a_1}+\ket{a_2,a_2}\right)\nonumber \\
               &=&\frac{1}{\sqrt{3}}\left(\ket{{a}_{0}',{a}_{0}'}+
               \ket{{a}_{1}',{a}_{2}'}+\ket{{a}_{2}',{a}_{1}'}\right)
\end{eqnarray}
Notice that in the $A'$ basis in order for
the results, obtained by Alice and Bob, to be perfectly correlated  does 
not mean that they 
will find the same state! For example, 
if Alice finds the
state $\ket{{a}_{2}'}$, the state $\ket{{a}_{1}'}$ is the one which makes
Bob perfectly correlated with Alice.

In order to write down the Bell inequality in a simple way, it is 
convenient to assign values to the various states, this assignment is 
shown in the table below 
\begin{center}
\begin{tabular}{|l||l|l||l|l|l|} \hline
value & $A$ & $A'$ & $M_0$ & $M_1$ &$M_2$ \\ \hline \hline
0 & $\ket{a_0}$ &$\ket{{a}_{0}'}$ &$\ket{m_{00}}$& $\ket{m_{01}}$&
$\ket{m_{02}}$ \\\hline
1 & $\ket{a_1}$ &$\ket{{a}_{1}'}$ &$\ket{m_{11}}$& $\ket{m_{12}}$&
$\ket{m_{10}}$ \\\hline
2 & $\ket{a_2}$ &$\ket{{a}_{2}'}$ &$\ket{m_{22}}$& $\ket{m_{20}}$&
$\ket{m_{21}}$ \\\hline
\end{tabular}
\end{center}
Notice that the $\ket{m_{kl}}$ -states have been organized in three 
sets, so that the value assigned to a given state is given by the first 
index. Moreover this organization into the sets $M_0$, $M_1$ and $M_2$, 
simplifies the notation in what follows. However, it is important to 
remember that the states in each set are {\bf not} orthogonal, in other 
words they do not form three orthogonal bases.

Contrary to how Bell inequalities usually are presented, we here first 
present the quantum limit and only afterwards the local variable limit.
The Bell inequality is obtained as the sum of probabilities for when the 
results of the measurements on the two qutrits are correlated and from 
this sum subtract all the probabilities for when the results are not 
correlated, i.e.
\begin{eqnarray}
B_3&=&\sum p(\rm{results~correlated})\nonumber \\
&-&\sum 
p(\rm{results~not~correlated})\nonumber
\end{eqnarray}
Now suppose that Alice measures in the $A$-basis and Bob measures a
projector in the $M_0$ set. For this combination of measurements, there 
are the following contributions to the sum $B_3$:
\begin{eqnarray}
P(M_0=A)&= &p(m_{00}\bigcap a_0)+p(m_{11}\bigcap a_1)+p(m_{22}\bigcap 
a_2)\nonumber \\
&=& 
\frac{1}{2}+\frac{1}{2\sqrt{3}}\\
P(M_0\neq A)&=& p(m_{11}\bigcap {a}_{0})+
p(m_{22}\bigcap {a}_{0})+p(m_{00}\bigcap {a}_{1})+p(m_{22}\bigcap 
{a}_{1})\nonumber \\
             &+&p(m_{00}\bigcap {a}_{2})+p(m_{11}\bigcap 
{a}_{2})\nonumber\\
&=&\frac{1}{2}-\frac{1}{2\sqrt{3}}
\end{eqnarray}
where $P(M_0=A)$ should be read as: Bob measures a projector in $M_0$ 
and 
Alice measures $A$ and Bob obtains the value which is correlated with   
Alice's result - hence the correct value. On the other hand $P(M_0\neq A)$ 
means that Bob is not correlated with Alice, and hence obtain an error. 
The 
probability 
$p(m_{kl}\bigcap 
a_n)=p(m_{kl}|a_n)p(a_n)$ 
is the joint probability 
for obtaining both $\ket{a_n}$ 
and $\ket{m_{kl}}$. 

The same is the case if Alice measures in $A$ and Bob the projectors in 
the sets $M_1$ 
or 
$M_2$, and again if Alice measures $A'$ and Bob  the projectors in 
$M_0$. 
This 
gives the contribution, from the $M_1-A$ combination of measurements:
$P(M_1=A)
=p(m_{01}\bigcap a_0)+p(m_{12}\bigcap a_1)+p(m_{20}\bigcap 
a_2)
=\frac{1}{2}+\frac{1}{2\sqrt{3}} $ and
$P(M_1\neq A)
=\frac{1}{2}-\frac{1}{2\sqrt{3}} $. And from the $M_2-A$ combination:
$P(M_2=A)
=\frac{1}{2}+\frac{1}{2\sqrt{3}}$ and
$P(M_2\neq A )=
\frac{1}{2}-\frac{1}{2\sqrt{3}}$. And finally from the $M_0-A'$ 
combination:
$P(M_0=A')=
\frac{1}{2}+\frac{1}{2\sqrt{3}}$ and
$P(M_0\neq A')
=\frac{1}{2}-\frac{1}{2\sqrt{3}}$. 

Consider now the case where Alice measures in $A'$ and Bob measures the 
states in the set  $M_1$. In this case Bob consistently finds a value 
which is two higher (modulus 
3) than the one which correlates him with  Alice. To see this, assume for 
example that Bob has the state $\ket{{a}_{0}'}$ which is assigned the 
value $0$. But the state which gives the correct identification of 
this state is $\ket{{m}_{20}}$, which is assigned the value $2$. Similar 
for the other states, which leads to
$P(M_1=A'+2) 
=p(m_{20}\bigcap 
{a}_{0}')+p(m_{01}\bigcap {a}_{1}')+p(m_{12}\bigcap {a}_{2}') 
=\frac{1}{2}+\frac{1}{2\sqrt{3}}$ and 
$P(M_1\neq A'+2)
=\frac{1}{2}-\frac{1}{2\sqrt{3}}$

Whereas if Alice measures in $A'$ and Bob the states in $M_2$, he 
consistently finds a value which is 1 higher than the value which 
correlates him with  
Alice, i.e.
$P(M_2=A'+1)
=\frac{1}{2}+\frac{1}{2\sqrt{3}}$ and
$P(M_2\neq A'+1)
=\frac{1}{2}-\frac{1}{2\sqrt{3}}$.

The sum $B_3$ may now be written and evaluated:
\begin{eqnarray}
B_3&=&P(M_0=A)- P(M_0\neq A) \nonumber \\
&+&P(M_1=A)- P(M_1\neq A) \nonumber \\  
&+&P(M_2=A)- P(M_2\neq A) \nonumber\\
&+&P(M_0=A')- P(M_0\neq A') \nonumber \\
&+&P(M_1=A'+2)- P(M_1\neq A'+2) \nonumber\\
&+&P(M_2=A'+1)- P(M_2\neq A'+1) \nonumber \\
&=& 2\sqrt{3}
\end{eqnarray}
The quantity $B_3$ is a sum of joint probabilities and if written in full 
it consists of 54 terms. 
A local variable model which tries to attribute definite values to 
the observables will reach a maximum value of 2. This may be checked 
numerically, but it can also be argued as is done in the following.

Since $a_0$, $a_1$ and $a_2$ are measured simultaneously in a single
measurement of the basis $A$, only one of them can come out true in local 
variable model. The same
is the case for the ${a}_{0}'$, ${a}_{1}'$ and ${a}_{2}'$, which are
measured as the basis $A'$. This means that, for example, if $a_0$ is
true, meaning that a measurement of $A$ will result in the outcome $a_0$,
then all probabilities involving $a_1$ and $a_2$ must be zero. It is
different for the $\ket{m_{ij}}$ states since they are measured
independently and hence they may all be true at the same time in a local 
variable model.

Now assume that according to a local variable model ${a}_{i}$ and 
${a}_{j}'$ are true, at the same time in principle all the states $m_{kl}$ 
could be true too. 
The question is now what will be the contributions 
from the various $m-$states. There are several possibilities. The state 
$m_{kl}$ where both indices are different will only give negative 
contribution to the 
sum $B_3$, since it fails to make the correct identification of any of 
the two basis states, and therefore only give rise to errors, i.e. 
contribute with $-2$. Whereas a state $m_{il}$ 
or $m_{kj}$, where one index is 
correct, will lead to a correct identification of one of the basis states, 
but since it fails to correctly identify the other, the net result is that 
these $m-$states give no contribution to the total sum.

The only state which 
will 
gives a positive contribution to the total sum is  $m_{ij}$ which 
identifies correctly both  ${a}_{i}$ and ${a}_{j}'$, hence give 
a 
contribution of $+2$. From this it is seen that the maximum value 
according to local variables is 2, i.e.
\beq
B_3\leq 2
\label{eq:bell}
\eeq
However, we have already seen that quantum 
mechanically it is possible to violate this limit. Quantum mechanically 
the limit is $2\sqrt{3}$.  

It has been checked 
numerically that $2\sqrt{3}$ is indeed  the 
maximal quantum mechanical limit for this sum of probabilities
and that the quantum mechanical maximum is indeed reached for the maximally entangled state.
Moreover it has also been shown, using "polytope software" \cite{bb,poly},
that inequality (\ref{eq:bell}) is optimal for the measurement settings 
which we have presented here.

\section{Extension to arbitrary dimension $N$}
In this section we show how to construct the inequality in any 
dimension $N$. We again assume that Alice and Bob share many 
maximally entangled states. 

Consider again the two bases $A$ and $A'$, 
where the first 
is the computational basis and the second the Fourier transformed, each 
one now containing $N$ basis vectors. Since the two bases are mutually 
unbiased, the distance between any state from one basis to any state in 
the other basis, is $\cos(\theta)=1/\sqrt{N}$. 
The intermediate states may be 
constructed in exactly the same way as in three dimension, which means by 
forming all pairs of states from the two bases. Since the intermediate 
states (there are $N^2$) are defined as the ones lying exactly 
between a pair of states, 
which means that the distance from the intermediate state $\ket{{m}_{ij}}$ 
to the 
states $\ket{{a}_{i}}$ and $\ket{{a}_{j}'}$ is  
$\cos(\theta/2)=\sqrt{{\frac{1}{2}+\frac{1}{2\sqrt{N}}}}$. Which means 
that 
the probability of correct identification is 
\beq
p({m}_{ij}|{a}_{i})=p({m}_{ij}|{a}_{j}')=\frac{1}{2}+\frac{1}{2\sqrt{N}}
\eeq
and the probability of an error is 
\beq
p({m}_{ij}|{a}_{k})=p({m}_{ij}|{a}_{l}')=\frac{1}{N-1}
\left(\frac{1}{2}-\frac{1}{2\sqrt{N}}\right).   
\eeq
As in the case for two qutrits, it is convenient to assign values to the 
various states. In the table below is shown the values and the 
organization of the states into the sets $M_0$-$M_{N-1}$:
\begin{center}
\begin{tabular}{|l||l|l||l|l|l|l|}\hline
value & $A$ & $A'$ & $M_0$ & $M_1$ & $\cdots$  & $M_{N-1}$\\ 
\hline \hline
0 & $\ket{a_0}$ &$\ket{{a_0}'}$ &$\ket{m_{00}}$& $\ket{m_{01}}$&$\cdots$ 
&
$\ket{m_{0,N-1}}$ \\\hline
1 & $\ket{a_1}$ &$\ket{{a_1}'}$ &$\ket{m_{11}}$& $\ket{m_{12}}$ 
&$\cdots$&
$\ket{m_{10}}$ \\\hline
$\cdots$&$\cdots$&$\cdots$&$\cdots$&$\cdots$&$\cdots$&$\cdots$\\ 
\hline
$N-1$ & $\ket{a_{N-1}}$ &$\ket{{a_{N-1}}'}$ &$\ket{m_{N-1,N-1}}$& 
$\ket{m_{N-1,0}}$&$\cdots$&
$\ket{m_{N-1,N-2}}$ \\\hline
\end{tabular}
\end{center}
Keeping the same notation as in the previous sections the Bell inequality 
for any dimension may be written
\begin{eqnarray}
B_N&= &\sum_{i=0}^{N-1}P(M_i=A) -\sum_{i=0}^{N-1}P(M_i\neq A)\nonumber \\
&+&\sum_{i=0}^{N-1}P(M_i=A'+N-i) -\sum_{i=0}^{N-1}P(M_i\neq 
A'+N-i)\nonumber\\
&\leq& 2
\end{eqnarray}
The local variable limit can again be argued as for the case of two 
qutrits, namely the only $m-$state which will give a positive contribution 
is the one where both indices are the same as for the true basis states. 
Any other $m-$state will either give a negative contribution to the 
total sum or 
no contribution at all. 

If it is written out in full it consists of 
$2N\times N^2$ 
terms. Since again the inequality is the sum of all the correct guesses, 
subtracting all the wrong guesses, the quantum mechanical limit is found 
to be
\beq
{\rm 
QM}=2N
\left(\left(\frac{1}{2}+\frac{1}{2\sqrt{N}}\right)- 
\left(\frac{1}{2}-\frac{1}{2\sqrt{N}}\right)\right)=2\sqrt{N}
\eeq
Hence we have obtained an inequality where the violation increases with 
the square-root of the dimension.

It is important to realize that in the case $N=2$ the inequality $B_2$, is 
the famous CHSH-Inequality. In this case the two bases can be taken as 
the $z$-basis and the $x$-basis of a spin-$1/2$. If considering these two 
bases as axes on a great circle of the sphere, the intermediate states 
are the ones which lies at $\pm 45$ degrees. 
Notice that in this case the two sets of 
intermediate states $M_0$ and $M_1$ actually do form two orthogonal bases. 

\section{Resistance to noise}
In the resent papers on Bell inequalities, the  strength of the 
inequality has been 
measured in terms of it's resistance to noise. The question is how much 
noise 
can be added to the maximally entangled state, $\ket{\psi}$, and still 
obtain a Bell 
violation. The more noise which can be added the better, since this means 
that the inequality 
is robust.

Until recently the noise considered was the uncolored noise, which means 
that the quantum state becomes
\begin{eqnarray}
\rho_{mix} = {\lambda}_{mix} \ket{\psi}\bra{\psi} + (1- 
{\lambda}_{mix})\frac{\openone}{N}
\end{eqnarray}
The Bell inequality we have presented here reaches the classical limit, 
$B_N=2$ for ${\lambda}_{mix}^{B_N}= \frac{N-1}{N+\sqrt{N}-2}$. For $N=3$ 
this 
is ${\lambda}_{mix}^{B_3}= \frac{2}{1+\sqrt{3}}\simeq 0.73$. In 
comparison, 
the inequality presented by Collins, Gisin, Linden, Massar and Popescu 
(CGLMP) 
\cite{bf} is more robust to this kind of noise 
since they find a violation until ${\lambda}_{mix}^{CGLMP}\simeq 0.69$. 

However it has recently been argued that the use of uncolored noise in 
this 
measure  lead to problems \cite{cp,bm}. At the same time a new 
idea was introduces, namely instead of mixing the maximally entangled 
state with the maximally mixed state, to mix it with the 
closest separable 
state, i.e.
\begin{eqnarray}
\rho_{cs} = {\lambda}_{sep} \ket{\psi}\bra{\psi} + (1-
{\lambda}_{sep}){\rho}_{sep}
\end{eqnarray}
where ${\rho}_{sep}=\frac{1}{N}\sum_{i=0}^{N-1}\ket{a_i,a_i}\bra{a_i,a_i}$ 
\cite{PV}. Making use of $\rho_{sep}$ leads to ${\lambda}_{sep}^{B_N}= 
\frac{N-\sqrt{N}}{N+\sqrt{N}-2}$, which for $N=3$ is 
${\lambda}_{sep}^{B_3}=
\frac{3-\sqrt{3}}{1+\sqrt{3}}\simeq 0.46$. Whereas the CGLMP inequality 
again has ${\lambda}_{sep}^{CGLMP}\simeq 0.69$. Which means that the 
inequality we introduce here is much more resistant to this kind of noise.

It should  however be stressed that the same measurement settings have 
been 
used in 
both evaluation of $\lambda$, and that the CGLMP inequality has been 
optimized to be resistant to the uncolored noise. It is nevertheless 
interesting to see how robustness of the $B_N$ inequality change depending
on the different noise added to the system.

\section{Conclusion}
We have presented a Bell inequality for quNits. 
The 
classical 
limit for this particular sum of joint probabilities is $2$ - independent 
of the dimension. 
Whereas quantum mechanically it is 
possible to obtain a violation which increases with the square-root of 
the dimension, namely    
$2\sqrt{N}$. One of the interesting features of this inequality are the 
measurements which lead to the maximal violation. 
On Alice's side we have 
the 
usual  two    
standard  measurements of two mutually unbiased bases, but on Bob's 
side there is the choice of  $N^2$ binary measurements. These 
measurements, which are represented by non-orthogonal projectors,  
correspond
to the intermediate states of the two bases used by Alice.   
Intermediate states are known from intercept/resend 
eavesdropping in quantum cryptography.

It was the observation that for qubits, the intermediate states are both 
used in quantum cryptography and as the maximal settings for the 
CHSH-inequality which lead us to this construction of Bell inequalities in 
higher dimension.
For $N=2$, we therefore also recover the familiar CHSH-inequality. 

This  inequality further more has the advantage of being easily derived in 
any dimension. This is again due to the measurement settings. Since both 
the choice of measurements on Alice's side and on Bob's side are easily 
generalized to arbitrary dimension, so is the inequality. 

Until recently the strenght of an inequality has been measured in terms of 
its resistance to uncolored noise. The inequality we present here is less 
resistant to this kind of noise than others. On the other hand it was 
recently argued \cite{bm} that the use of uncolored noise leads to 
problems. Instead it was suggested to mix the maximally entangled state 
with the closest separable state. Using this kind of noise we have shown 
that the inequality presented here is much more robust than the CGLMP 
inequality.

Finally, it should also be mentioned that, in contrast to several other     
inequalities which have been
presented recently, we have maximal violation for the
maximally entangled state. This means that the inequality which we
present here may be used as a measure of entanglement.

\section*{Acknowledgments}
H.B.-P. is supported by the Danish National Science
Research Council (grant no. 9601645) and the Swiss NCCR "Quantum 
Photonics".

\end{document}